\documentclass[adp,fleqn]{w-art}
\usepackage{times}
\usepackage{w-thm}
\usepackage[]{graphicx}
\chardef\bslash=`\\ 

\begin{document}
\DOIsuffix{theDOIsuffix}
\Volume{12}
\Issue{1}
\Copyrightissue{01}
\Month{01}
\Year{2003}
\pagespan{1}{}
\Receiveddate{22 October 2008 by U. Eckern}



\title[Superfluid to Mott-insulator transition]{Superfluid to Mott-insulator transition
in  an anizotropic\\ two--dimensional optical lattice}


\author[T. P. Polak]{T. P. Polak\footnote{Corresponding
     author{\quad}E-mail: {\sf tppolak@amu.edu.pl	}}\inst{1}} 
\address[\inst{1}]{Adam Mickiewicz University of Pozna\'{n}, Faculty of Physics, Umultowska
85, 61-614 Pozna\'{n}, Poland}
\author[T. K. Kope\'c]{T. K. Kope\'c\footnote{E-mail: {\sf kopec@int.pan.wroc.pl} 
        }\inst{2}}
\address[\inst{2}]{Institute for Low Temperatures and Structure Research, Polish Academy
of Sciences, POB 1410, 50-950 Wroclaw 2, Poland}


\begin{abstract}
We study the  superfluid to Mott-insulator
transition of bosons in an optical anizotropic lattice by employing
the Bose-Hubbard model living on a two-dimensional lattice with anizotropy
parameter $\kappa$. The compressible superfluid state and incompressible
Mott-insulator (MI) lobes are efficiently described analytically,
using the quantum U(1) rotor approach. The ground
state phase diagram showing the evolution of the MI lobes is quantified
for arbitrary values of $\kappa$, corresponding to various kind of
lattices: from square, through rectangular to almost one-dimensional.
\end{abstract}
\maketitle                   





\section{Introduction}
\label{sect1}
The type of order that a physical system can possess is utterly affected
by its dimensionality. In two-dimensional ($2D$) systems with a continuous
symmetry  long-range order is destroyed  by  fluctuations
at a finite temperature \cite{mermin}. However, the competition between
ground states in $T=0$ can lead to a zero-temperature phase transition,
driven solely by quantum mechanical fluctuations. In this context unconventional
behavior in low-dimensional systems was intensively studied in the past
years \cite{sachdev}.

Since the experimental realization of Bose-Einstein condensation \cite{anderson}
many properties of it have been elucidated \cite{jaksch,greiner}.
Atomic gases allow clean and controlled observation of  variety physical
phenomena from condensed matter physics, e.g., the Berezinskii-Kosterlitz-Thouless
(BKT) phase transition with the emergence of  topological order \cite{hadzibabic}.
The merging of atomic and condensed matter physics has opened exciting
new perspectives for the creation of novel quantum states. Especially,
systems of cold atoms in optical lattices \cite{greiner1,gorlitz}
facilitate an experimental environment, where a rich variety of quantum
many-body models can be implemented in a wide range of spatial dimensions,
geometries, and particle interactions. Among these topics the emergence
of condensation and superfluid order in an optical lattice 
has been a major issue in recent years  \cite{mun,jo,oshikawa}.
In this context, the presence of the optical trapping structure offers
a unique way to increase (or decrease) the dimensionality of the system
in a clean experimental setup, thus providing a playground for studying
the effect of dimensional crossover on the quantum phase transition.
This can happen in optical lattices in which atoms can tunnel easily
along one spatial direction but not along the other one. This motivates
the analysis of very interesting physics, e.g., of an anisotropic array of coupled
one-dimensional ($1D$) Bose gases. Here, the coupling is provided
by the intersite tunneling that can be made variable  by adjusting
the optical lattice potential. A one-dimensional situation is created by
suppressing tunneling in two directions by using two standing waves
with very high laser intensities that control the barrier between
the lattice sites.

The aim of this work is to study the  superfluid (SF) to Mott-insulator (MI) transition by means of the Bose-Hubbard model in $2D$ optical lattices with
variable lattice anizotropy parameter. In order  to quantify the
evolution of the ground state phase diagram,   various
kind of lattices, from square, through rectangular to almost one--dimensional will be considered.
 To this we have adopted a theoretical
approach to the strongly interacting fermions \cite{kopec} in the
Bose-Hubbard model in a way to include particle number fluctuations
effects and make the qualitative phase diagrams more quantitative 
\cite{polak}. The key point of our approach is to consider the representation
of strongly interacting bosons as particles with attached {}``flux
tubes''. This introduces a U(1) phase variable,
which acquires dynamic significance from the boson-boson interaction.
In the present work we investigate the dimensional crossover from two- to almost $1D$
lattices and obtain the  ground state phase diagram.

\section{The model and method}
\label{sect2}
We start with the generic model for the Mott-insulator transition, namely 
the Bose-Hubbard model \cite{fisher}
\begin{equation}
\mathcal{H}=\frac{U}{2}\sum_{i}n_{i}\left(n_{i}-1\right)-\sum_{\left\langle i,j\right\rangle }t_{ij}a_{i}^{\dagger}a_{j}-\mu\sum_{i}n_{i},\label{hamiltonian1}
\end{equation}
where $a_{i}^{\dagger}$ and $a_{j}$ stand for the bosonic creation
and annihilation operators that obey the canonical commutation relations
$[a_{i},a_{j}^{\dagger}]=\delta_{ij}$, $n_{i}=a_{i}^{\dagger}a_{i}$
is the boson number operator on the site $i$, $U>0$ is the on-site
repulsion and the chemical potential $\mu$ controls the number of
bosons. Here, $\left\langle i,j\right\rangle $ identifies summation
over the nearest-neighbor sites. Furthermore, $t_{ij}$ is the hopping
matrix element with dispersion 
\begin{equation}
t\left(\mathbf{k}\right)=2t\left(\cos k_{x}+\kappa\cos k_{y}\right),
\end{equation}
where $\kappa$ is the anizotropy parameter and $t$ sets the kinetic
energy scale for bosons. By varying quantity $\kappa$ between zero
($1D$) and one ($2D$), different anizotropic rectangular lattices
emerge. 

We write the partition function of the system
\begin{equation}
\mathcal{Z}  =  \int\left[\mathcal{D}\bar{a}\mathcal{D}a\right]\exp\left[-\int_{0}^{\beta}d\tau\mathcal{H\left(\tau\right)}-\sum_{i}\int_{0}^{\beta}d\tau\bar{a}_{i}\left(\tau\right)\frac{\partial}{\partial\tau}a_{i}\left(\tau\right)\right]
\end{equation}
using the bosonic path-integral over the complex fields $a_{i}\left(\tau\right)$
depending on the {}``imaginary time'' $0\leq\tau\leq\beta\equiv1/k_{\mathrm{B}}T$
with $T$ being the temperature. We decouple the interaction term
in Eq. (\ref{hamiltonian1}) by a Gaussian integration over the auxiliary
scalar potential fields
\begin{equation}
 V_{i}\left(\tau\right)=V_{i0}+V_{i}'\left(\tau\right)
\end{equation}
with static $\beta V_{i0}=V_{i}\left(\omega_{\nu}=0\right)$ and periodic
part 
\begin{equation}
V'_{i}\left(\tau\right)=\beta^{-1}\sum_{\nu=1}^{+\infty}V_{i}\left(\omega_{\nu}\right)\exp\left(i\omega_{\nu}\tau\right)+\mathrm{c.c},
\end{equation}
where $\omega_{\nu}=2\pi\nu/\beta$ ($\nu=0,\pm1,\pm2,...$) is the
Bose-Matsubara frequency. Periodic part $V'_{i}\left(\tau\right)\equiv V'_{i}\left(\tau+\beta\right)$
couples to the local particle number through the Josephson-like relation
\begin{equation}
\dot{\phi}_{i}\left(\tau\right)=V'_{i}\left(\tau\right)
\end{equation}
 where $\dot{\phi}_{i}\left(\tau\right)\equiv\partial\phi_{i}\left(\tau\right)/\partial\tau$.
The quantity $\phi\left(\tau\right)$ is the phase field satisfies
the periodicity condition $\phi_{i}\left(\beta\right)=\phi_{i}\left(0\right)$
as a consequence of the periodic properties of the $V'_{i}\left(\tau\right)$
field. Further, we perform the local gauge transformation to the new
bosonic variables 
\begin{equation}
a_{i}\left(\tau\right)=b_{i}\left(\tau\right)\exp\left[i\phi_{i}\left(\tau\right)\right]
\end{equation}
 that removes the imaginary term $-i\int_{0}^{\beta}d\tau\dot{\phi_{i}}\left(\tau\right)n_{i}\left(\tau\right)$
from all the Fourier modes. From the above we deduce bosons have a
composite nature made of bosonic part $b_{i}\left(\tau\right)$ and
attached {}``flux'' $\exp\left[i\phi_{i}\left(\tau\right)\right]$.
Note that a similar method was used in 
a functional-integral formulation to treat the quantum dynamics of a microscopic model of a Josephson junction, including the dissipative effects of quasiparticle tunneling \cite{eckern}.
Next, we parameterize the boson fields
\begin{equation}
 b_{i}\left(\tau\right)=b_{0}+b_{i}^{'}\left(\tau\right)
\end{equation}
and incorporate fully our in calculations the phase fluctuations governed
by the gauge $\mathrm{U}\left(1\right)$ group and drop corrections
to the amplitude by assuming $b_{i}\left(\tau\right)=b_{0}$, which
was proven to be justified in the large $U/t$ limit we are interested
in \cite{polak,kampf}. By integrating out the auxiliary static field
$V_{i0}$ we calculate the partition function $\mathcal{Z}=\int\left[\mathcal{D}\phi\right]\exp\left\{ -\mathcal{S}_{\mathrm{ph}}\left[\phi\right]\right\} ,$
with an effective action expressed in phase-only terms
\begin{equation}
\mathcal{S}_{\mathrm{ph}}\left[\phi\right]  =  \int_{0}^{\beta}d\tau\left\{ \sum_{i}\left[\frac{1}{2U}\dot{\phi_{i}^{2}}\left(\tau\right)+\frac{1}{i}\frac{\bar{\mu}}{U}\dot{\phi_{i}}\left(\tau\right)\right]-\sum_{\left\langle i,j\right\rangle }e^{i\phi_{i}\left(\tau\right)}J_{ij}e^{-i\phi_{j}\left(\tau\right)}\right\} .\label{action only phase}
\end{equation}
We note that the phase action  for  the Bose-Hubbard model is closely related to the standard model of Josephson junction arrays, which contains the charging energy as well as the Josephson coupling
\cite{jja1, jja2}.
\begin{vchfigure}
\includegraphics[%
  scale=0.45]{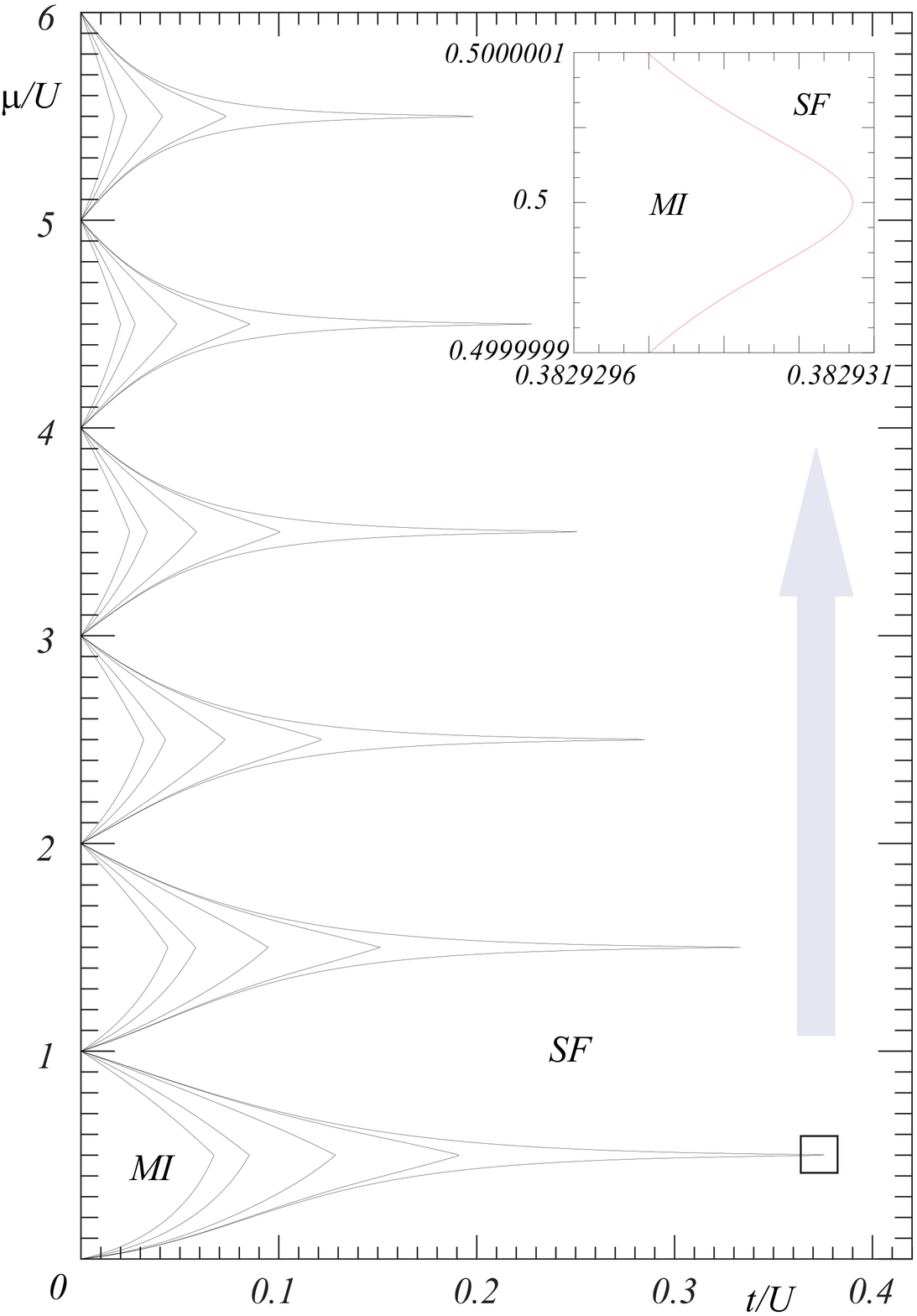}
\vchcaption{(online colour) The phase boundary between the Mott insulating and
superfluid phases for the anizotropic lattice with $\kappa=1,0.5,10^{-1},10^{-2},10^{-5}$
from left to right. Inset shows the tip of the first lobe ($n_{B}=1$)
for $\kappa=10^{-5}$.\label{phase}}
\end{vchfigure}
The phase stiffness coefficient is given by $J_{ij}=b_{0}^{2}t_{ij}$,
where the value
\begin{equation}
b_{0}^{2}=\frac{1}{U}\left(\frac{1}{N}\sum_{\left\langle i,j\right\rangle }t_{ij}+\bar{\mu}\right)
\end{equation}
 is obtained from minimalization of the Hamiltonian $\partial\mathcal{H}\left(b_{0}\right)/\partial b_{0}=0$;
$\bar{\mu}/U=\mu/U+1/2$ is the shifted reduced chemical potential.
The total time derivative Berry phase imaginary term in Eq. (\ref{action only phase})
is nonzero due to phase field configurations with 
\begin{equation}
\phi_{i}\left(\beta\right)-\phi_{i}\left(0\right)=2\pi m_{i}
\end{equation}
where, $m_{i}=0,\pm1,\pm2...$. Therefore, we concentrate on closed paths
in the imaginary time $\left(0,\beta\right)$ labelled by the integer
winding numbers $m_{i}$. The path-integral 
\begin{equation}
\int\left[\mathcal{D}\phi\right]...\equiv
\sum_{\left\{ m_{i}\right\} }\int_{0}^{2\pi}
\left[\mathcal{D}\phi\left(0\right)\right]\int_{_{\phi_{i}\left(0\right)}}^{\phi_{i}\left(\tau\right)+2\pi m_{i}}\left[\mathcal{D}\phi\left(\tau\right)\right]...,
\end{equation}
includes a summation over $m_{i}$ and in each topological sector
the integration goes over the gauge potentials.
To proceed, we replace the phase degrees of freedom by the unimodular
scalar complex field $\psi_{i}$ which satisfies the quantum periodic
boundary condition $\psi_{i}\left(\beta\right)=\psi_{i}\left(0\right)$.
This can be conveniently done using the Fadeev-Popov method with Dirac
delta functional resolution of unity \cite{kopec1}, where we take
$\psi_{i}$ as continuous but constrained (on the average) variable
to have the unimodular value
\begin{equation}
1  =  \int\left[\mathcal{D}\psi\mathcal{D}\bar{\psi}\right]\delta\left(\sum_{i}\left|\psi_{i}\left(\tau\right)\right|^{2}-N\right) \prod_{i}\delta\left(\psi_{i}-e^{i\phi_{i}\left(\tau\right)}\right)\delta\left(\bar{\psi}_{i}-e^{-i\phi_{i}\left(\tau\right)}\right).\label{popov}
\end{equation}
 Introducing the Lagrange multiplier $\lambda$, which adds the quadratic
terms (in the $\psi_{i}$ fields) to the action Eq. (\ref{action only phase}),
we can solve the constraint. Using such description is justified by
the definition of the order parameter
 \begin{equation}
\Psi_{B}\equiv\left\langle a_{i}\left(\tau\right)\right\rangle =\left\langle b_{i}\left(\tau\right)\exp\left[i\phi_{i}\left(\tau\right)\right]\right\rangle =b_{0}\psi_{B},
\end{equation}
 which non-vanishing value signals a bosonic condensation (we identify
it as superfluid state). Note that a nonzero value of the amplitude
$b_{0}$ is not sufficient for superfluidity. To achieve this, also
the phase variables, must become stiff and coherent, which implies
$\psi_{B}\neq0$. 

\section{Phase diagram}

The partition function is written in the form
\begin{eqnarray}
\mathcal{Z} & = & \int_{-i\infty}^{+i\infty}\left[\frac{\mathcal{D}\lambda}{2\pi i}\right]e^{-N\beta\mathcal{F}\left(\lambda\right)},
\end{eqnarray}
 with the free energy density $\mathcal{F}=-\ln\mathcal{Z}/\beta N$
given by:
\begin{equation}
\mathcal{F} = -\lambda-\frac{1}{N\beta}\ln\int\left[\mathcal{D}\psi\mathcal{D}\bar{\psi}\right]\exp\left\{ \sum_{i,j}\int_{0}^{\beta}d\tau d\tau'\left[\left(J\mathcal{I}_{ij}+\lambda\delta_{ij}\right)\delta\left(\tau-\tau'\right) \right ]
-\mathcal{\gamma}_{ij}\psi_{i}\bar{\psi}_{j}\right\}
 ,\label{free energy}
\end{equation}
 where $\mathcal{I}_{ij}=1$ if $i,j$ are the nearest neighbors and
equals zero otherwise,
\begin{equation}
 \gamma_{ij}\left(\tau,\tau'\right)=\left\langle
 \exp\left\{ -i\left[\phi_{i}\left(\tau\right)
-\phi_{j}\left(\tau'\right)\right]\right\} \right\rangle 
\end{equation}
is the two-point phase correlator associated with the order parameter
field, where $\left\langle ...\right\rangle $ is the averaging with
respect to the action in Eq. (\ref{action only phase}). The action
with the topological contribution, after Fourier transform, we write
as
 \begin{equation}
\mathcal{S}_{\mathrm{eff}}\left[\psi,\bar{\psi}\right]
=\frac{1}{N\beta}\sum_{\mathbf{k},\nu}
\bar{\psi}_{\mathbf{k},\nu}\mathrm{\Gamma}_{\mathbf{k}}^{-1}
\left(\omega_{\nu}\right)\psi_{\mathbf{k},\nu},
\end{equation}
where 
\begin{equation}
\mathrm{\Gamma}_{\mathbf{k}}^{-1}\left(\omega_{\nu}\right)
=\lambda-J_{\mathbf{k}}+\gamma^{-1}\left(\omega_{\nu}\right)
\end{equation}
is the inverse of the propagator. The final form of the correlator,
after Fourier transform, can be written as:
\begin{equation}
\gamma\left(\omega_{\nu}\right)=
\frac{1}{\mathcal{Z}_{0}}
\frac{4}{U}\sum_{m}
\frac{\exp\left[-\frac{U\beta}{2}\left(m+\frac{\bar{\mu}}{U}\right)^{2}\right]}
{1-4\left(m+\frac{\bar{\mu}}{U}-i\frac{\omega_{\nu}}{U}\right)^{2}},
\label{correlator}
\end{equation}
where 
\begin{equation}
\mathcal{Z}_{0}=\sum_{m}\exp\left[-U\beta\left(m+\bar{\mu}/U\right)^{2}/2\right]
\end{equation}
is the partition function for the set of quantum rotors. The form
of Eq. (\ref{correlator}) assures the periodicity in the imaginary
time with respect to $\mu/U+1/2$ which emphasizes the special role
of its integer values (see Fig. \ref{phase}). Within the phase coherent
state the order parameter $\psi_{B}$ is evaluated in the thermodynamic
limit $N\rightarrow\infty$ by the saddle point method $\delta\mathcal{F}/\delta\lambda=0$
takes the form:
\begin{equation}
1-\psi_{B}^{2}=\frac{1}{N\beta}\sum_{\mathbf{k},\nu}\frac{1}{\lambda-J_{\mathbf{k}}+\gamma^{-1}\left(\omega_{\nu}\right)}.\label{critical line}
\end{equation}
\begin{vchfigure}
\includegraphics[%
  scale=0.41]{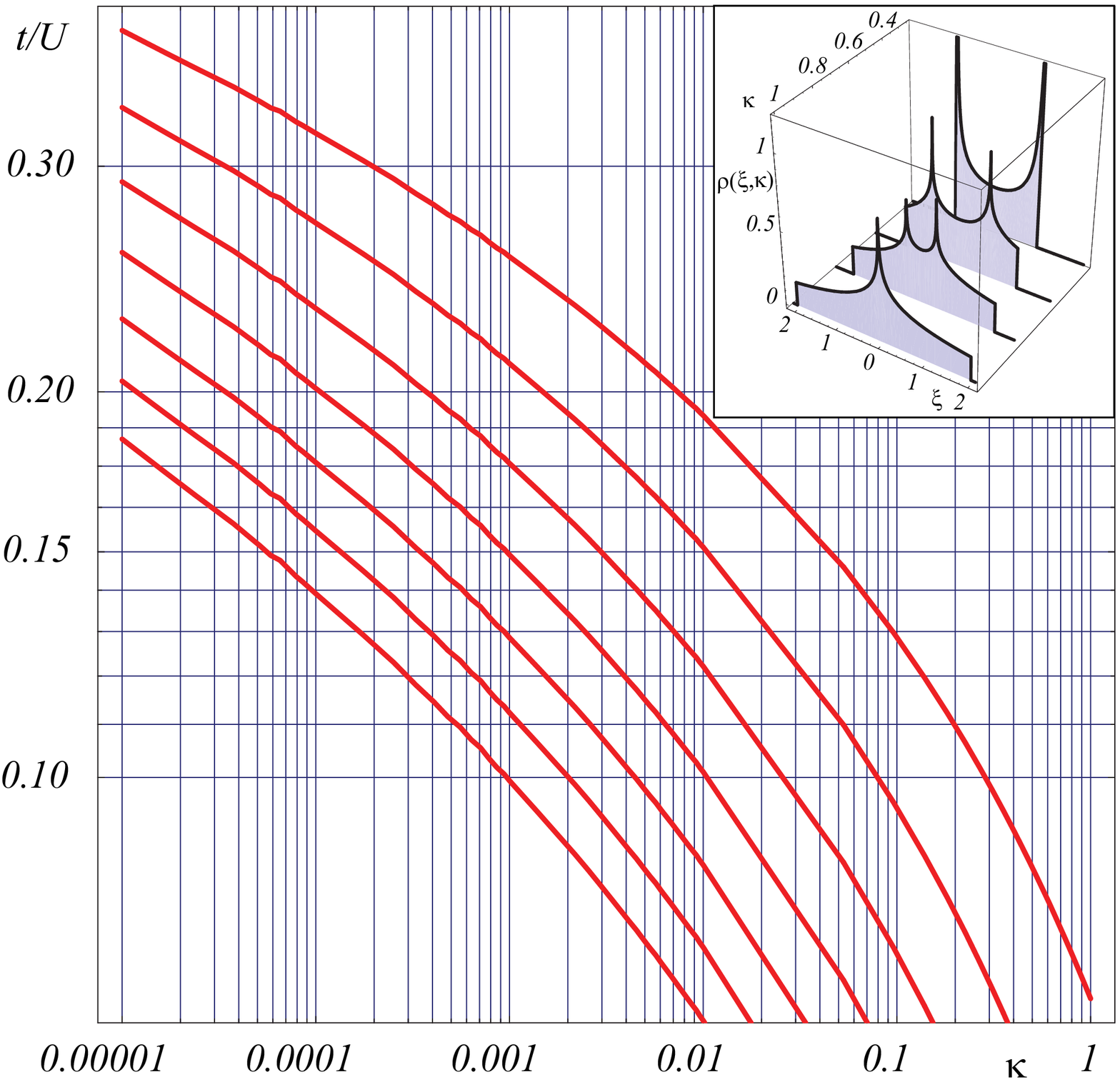}
\vchcaption{(online colour) The position of the tips of the lobes as a function
of the anizotropy parameter $\kappa$ for $\mu/U=0.5,1.5,2.5,3.5,4.5,5.5,6.5$
from right to left in double logarithmic scale. Inset shows the evolution
of the density of states $\rho\left(\xi,\kappa\right)$ from square
lattice ($\kappa=1$) through rectangular ($\kappa=0.33,0.66$) to
one-dimensional ($\kappa=0$).\label{tips}}
\end{vchfigure}
The phase boundary is determined by the divergence of the order parameter
susceptibility $\Gamma_{\mathbf{k}=0}\left(\omega_{\nu=0}\right)=0$,
which determines the critical value of the Lagrange parameter $\lambda=\lambda_{0}$
that stays constant in the whole ordered phase. We introduce the density
of states 
\begin{equation}
\rho\left(\xi,\kappa\right)=\frac{1}{N}\sum_{\mathbf{k}}\delta\left[\xi-t\left(\mathbf{k}\right)/t\right]
\end{equation}
for rectangular anizotropic lattice (see inset Fig. \ref{tips}):
\begin{eqnarray}
\rho\left(\xi,\kappa\right) & = & \frac{1}{\pi^{2}\sqrt{\kappa}}\mathbf{K}\left(\sqrt{\frac{\left(1+\kappa\right)^{2}-\xi^{2}}{4\kappa}}\right)\nonumber \\
 & \times & \left[\Theta\left(\kappa-\left|\xi-1\right|\right)+\Theta\left(\kappa-\left|\xi+1\right|\right)\right]\nonumber \\
 & + & \frac{2\Theta\left(1-\kappa-\left|\xi\right|\right)}{\pi^{2}\sqrt{\left(1+\kappa\right)^{2}-\xi^{2}}}\mathbf{K}\left(\sqrt{\frac{4\kappa}{\left(1+\kappa\right)^{2}-\xi^{2}}}\right),\end{eqnarray}
 where $\Theta\left(x\right)$ is the unit step function and $\mathbf{K}\left(x\right)$
is the elliptic function of the first kind \cite{abramovitz}. With
help of the above and after summation over $\omega_{\nu}$, the superfluid
state order parameter becomes
\begin{equation}
1-\psi_{B}^{2}=\frac{1}{2}\int_{-\infty}^{+\infty}\frac{\rho\left(\xi,\kappa\right)d\xi}{\sqrt{\frac{J_{0}-t\xi}{U}+\upsilon^{2}\left(\frac{\mu}{U}\right)}}.\label{critical line final}
\end{equation}
 In Eq. (\ref{critical line final}) $\upsilon\left(\mu/U\right)=\mathrm{frac}\left(\mu/U\right)-1/2,$
where $\mathrm{frac}\left(x\right)=x-\left[x\right]$ is the fractional
part of the number and $\left[x\right]$ is the floor function which
gives the greatest integer less than or equal to $x$. 
The zero-temperature phase diagram of the model can be calculated
from Eq. (\ref{critical line final}). We recover the results for
pure $2D$ case (the lowest lobes in Fig. \ref{phase}) \cite{polak}.
Still there is a particle-hole asymmetry visible not in the position
of the maximum of the lobe - like in simple cubic lattice - but in
the shape of the curves. However, when we approach $1D$ case the
particle-hole symmetry is restored. In the context of the paper \cite{mukhopadhyay}
the existence of asymptotic restoration of the statistical particle-hole
symmetry in $1D$ dirty boson problem seems to be apparent. 
\begin{vchfigure}
\includegraphics[%
  scale=0.65]{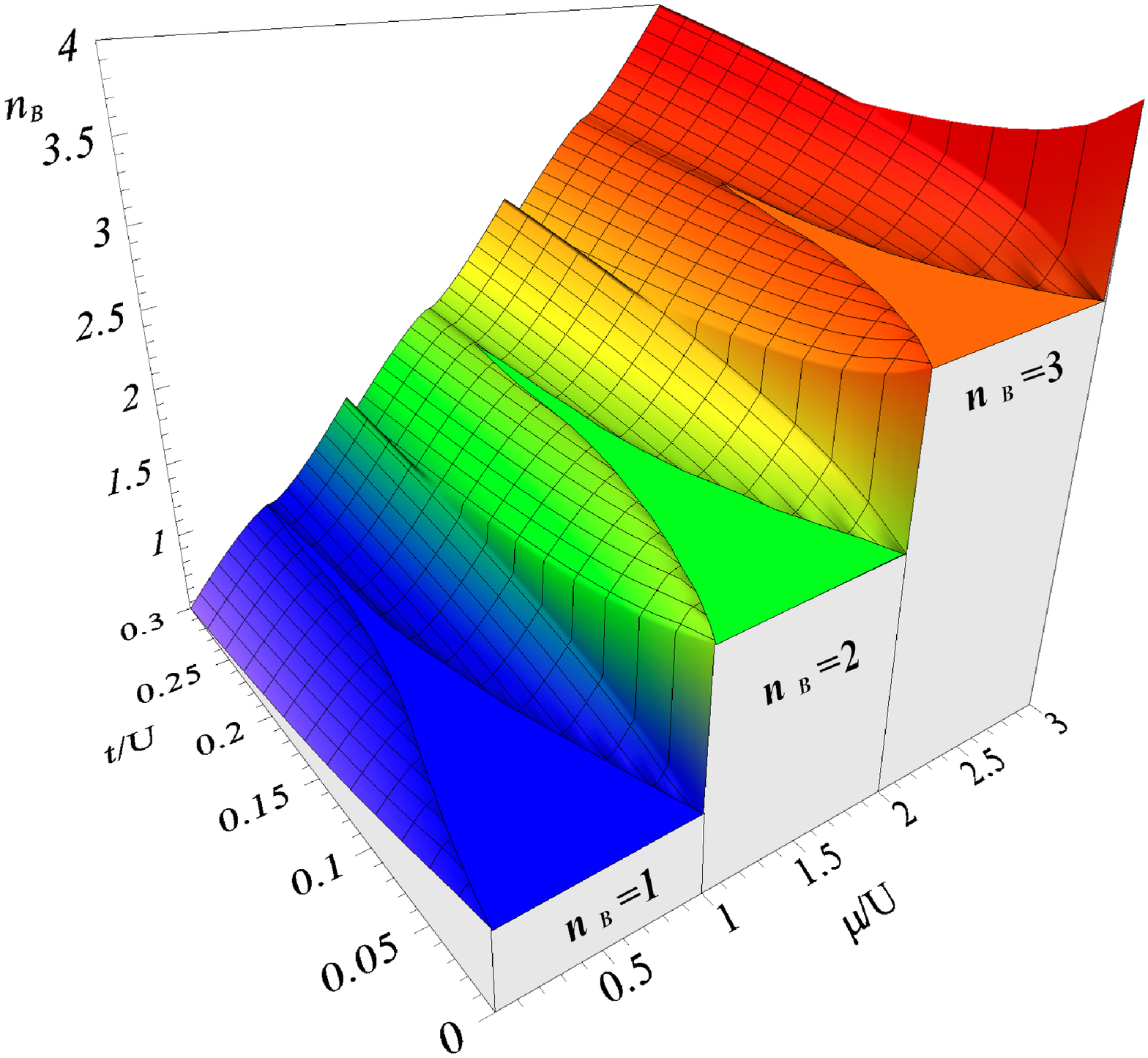}
\vchcaption{(online colour) Boson occupation number $n_{B}$ at $T=0$ for the
anizotropic lattice with $\kappa=10^{-5}$ as a function of chemical
potential $\mu/U$ and hopping $t/U$. The Mott insulator is found
within each lobe of integer boson density. Inside each of MI lobe
the integer occupation number $n_{B}$ is indicated.\label{nb}}
\end{vchfigure}
Next, we resort to the unimodular-field description and calculate
the effects of the fixed boson number 
\begin{equation}
n_{B}\equiv \frac{1}{N}\sum_{i}\left\langle \bar{a}_{i}\left(\tau\right)
a_{i}\left(\tau\right)\right\rangle 
\end{equation}
in the system. For the interacting problem, with the full phase action
Eq. (\ref{action only phase}) we get\begin{equation}
n_{B}=\left\{ \begin{array}{cc}
n_{B}\left(\lambda\right) & \textrm{within MI phase}\\
n_{B}\left(\lambda_{0}\right)-2\psi_{B}^{2}\upsilon\left(\frac{\mu}{U}\right) & \textrm{within SF phase}\end{array}\right.,\end{equation}
where $\psi_{B}$ is given by Eq. (\ref{critical line final}). In
the limit $T\rightarrow0$ an analytical solution of the total boson
density consists of the occupation number for neutral bosons $n_{b}$
and a contribution $\delta n_{b}$ from a fluctuating phase field:\begin{eqnarray}
n_{B}\left(\lambda\right) & = & \frac{\mu}{U}+\frac{1}{2}-\upsilon\left(\frac{\mu}{U}\right)\nonumber \\
 & - & \int_{-\infty}^{+\infty}d\xi\frac{\rho\left(\xi,\kappa\right)}{\sqrt{\frac{J_{0}-t\xi}{U}+\delta\lambda+\upsilon^{2}\left(\frac{\mu}{U}\right)}}\end{eqnarray}
with $\delta\lambda=\lambda-\lambda_{0}$. Here, the parameter $\lambda$
is self-consistently determined via Eq. (\ref{critical line}). Changes
in the anizotropy $\kappa$ provide the condition for emerging MI
(see Fig. \ref{tips}), with recognizable steps-like structure (Fig.
\ref{nb}). We cannot obtain pure $1D$ case since even in $T=0$
long-range order is destroyed by the quantum fluctuations.
Regarding the comparision of our method with the previous appraoches, e.g.,
\cite{fisher, kampf} we note that the qualitative
shape of the lobes resulting from our apprach is not the same for $2D$ and $3D$ cases, and
steeper for the two-dimensional system. 
Furthermore, we found \cite{polak} that our results are in good agreement  with the
recently published quantum Monte Carlo calculations on three-dimensional Bose--Hubbard
system \cite{capro}.

\section{Final remarks}

In conclusion, we have performed a study 
of the superfluid to Mott-insulator
transition of bosons in an optical anizotropic lattice by employing
the Bose-Hubbard model living on a two-dimensional lattice with anizotropy
parameter $\kappa$. The compressible superfluid state and incompressible
Mott-insulating  lobes are efficiently described analytically,
using the quantum U(1) rotor approach.
Our motivation comes from the fact that the development in experimental
techniques of trapping and controlling quantum gases  allow to
investigate such phenomena by offering a platform for the exploration
of highly non-trivial quantum phases and critical phenomena in dimensionality-tunable
systems. The technique used in
this paper can be easily extended to more general situations,
including, e.g., multi-species bosonic systems. Other generalizations
of the Bose--Hubbard model are of also possible by incoporating the
influence of disorder.

\begin{acknowledgement}
We thank R. Micnas for fruitful and stimulating discussions. One of
us (T.K.K) acknowledges the support by the Ministry of Education and
Science MEN under Grant No. 1 P03B 103 30 in the years 2006-2008.
\end{acknowledgement}

\end{document}